# Observation of hydrogen-induced Dzyaloshinskii–Moriya interaction and reversible switching of magnetic chirality

Gong Chen[1,2*], MacCallum Robertson[1], Markus Hoffmann[3], Colin Ophus[4], André L. Fernandes Cauduro[4], Roberto Lo Conte[5,6], Haifeng Ding[7], Roland Wiesendanger[6], Stefan Blügel[3], Andreas K. Schmid[4], Kai Liu[1,2,*]

[1] Physics Department, University of California, Davis, California 95616, USA

[2] Physics Department, Georgetown University, Washington, DC 20057, USA

[3] Peter Grünberg Institut and Institute for Advanced Simulation, Forschungszentrum Jülich and JARA, 52425 Jülich, Germany

[4] NCEM, Molecular Foundry, Lawrence Berkeley National Laboratory, Berkeley, California, 94720 USA

[5] Department of Materials Science & Engineering, University of California, Berkeley, California 95720, USA

[6] Department of Physics, University of Hamburg, D-20355 Hamburg, Germany

[7] National Laboratory of Solid State Microstructures, Department of Physics and Collaborative Innovation Center of Advanced Microstructures, Nanjing University, Nanjing 210093, P. R. China

* Correspondence should be addressed to gchenncem@gmail.com (G.C.); Kai.Liu@georgetown.edu (K.L.)

**ABSTRACT:** The Dzyaloshinskii–Moriya interaction (DMI) has drawn great attention as it stabilizes magnetic chirality, with important implications in fundamental and applied research. This antisymmetric exchange interaction is induced by the broken inversion symmetry at interfaces or in non-centrosymmetric lattices. Significant interfacial DMI was found often at magnetic / heavy-metal interfaces with large spin-orbit coupling. Recent studies have shown promise of induced DMI at interfaces involving light elements such as carbon (graphene) or oxygen. Here we report direct observation of induced DMI by chemisorption of the lightest element, hydrogen, on a ferromagnetic layer at room temperature, which is supported by density functional theory calculations. We further demonstrate a reversible chirality transition of the magnetic domain walls due to the induced DMI via hydrogen chemisorption/desorption. These results shed new light on the understanding of DMI in low atomic number materials and design of novel chiral spintronics and magneto-ionic devices.

## I. Introduction

The Dzyaloshinskii–Moriya interaction (DMI) is an antisymmetric exchange interaction, which can be induced in systems with broken inversion symmetry [1,2], such as the initially proposed $Fe_2O_3$ with weak ferrimagnetism, bulk materials with broken inversion symmetry like B20 compounds [3,4], or thin film systems [5,6]. The presence of the DMI may introduce various types of chiral magnetic structures, including helical spin spirals [3], helical/Bloch-type skyrmions [4,7], and magnetic bobbers [8] in bulk materials, or cycloidal spirals [5,9], Néel-type skyrmions [6,10-12] and Néel-type chiral domain walls [13-16] in thin films. The non-zero topological charge in chiral magnetic structures adds an exciting degree of freedom [7,17], which is fundamentally intriguing. It enables novel chiral spintronic applications with ultra-low energy consumption, e.g., racetrack memories or neuromorphic computing devices [10-12,18-21], where the ability to sensitively control the chirality, stability, or size of the spin texture is critical. For instance, in the limit of stable skryrmions, the radius of the skyrmion may be tailored by orders of magnitude with a fine tuning of the DMI strength [22]. Considering that many response quantities, e.g. magneto-resistive read-out, scale with the area of the skyrmion, thus controlling the DMI on a fine scale has important technological ramifications.

In thin films, it has been found experimentally that significant interfacial DMI can be induced when magnetic layers are adjacent to transition metals of large spin-orbit interaction [10-12], oxides [23,24], or light elements such as graphene [25,26] or oxygen [27]. Theoretical calculations based on the Fert-Levy model [28] reveal that the transition-metal induced DMI originates from the strong spin-orbit coupling combined with the wave function or charge asymmetry, which can be measured by the electric dipole moment [29]. The magnitude of the DMI is closely related to the band filling and hybridization between 5d and 3d orbitals near the Fermi level [30]. On the other hand, the light element induced DMI may be explained by the Rashba effect [25], and the charge asymmetry resulting from the charge transfer and hybridization of the band structure at the interface [31].

In this context manipulating surfaces and interfaces with the lightest element, hydrogen, is potentially a powerful path to control the DMI. Hydrogen *absorption* has been shown to significantly alter properties of magnetic materials, which permits tuning of interlayer exchange coupling [32], tailoring of magnetic anisotropy [33,34], or realization of magnetoelectric coupling in antiferromagnetic oxides [35]. Hsu et al. observed a hydrogenation-assisted formation of skyrmions in the Fe/Ir(111) system at 4.2 K, which is attributed to the modification of the Heisenberg exchange and DMI based on *ab initio* calculations [36]. A recent theory also predicts that hydrogen absorption on graphene alters the DMI at the graphene/Co interface [37].

On the other hand, few studies have distinguished hydrogen *chemisorption* induced effects on magnetic thin films, i.e., adsorption on the surface without penetrating the film. There have been reports on hydrogen adsorption induced effects on magnetic anisotropy [38,39]. However, direct and quantitative experimental confirmation of hydrogen-induced DMI is still lacking. Furthermore, Tan et al. have demonstrated $H^+$-based reversible magneto-ionic switching at room temperature where electric field-controlled hydrogen transport to the buried Co/GdO interface is used to toggle the perpendicular magnetic anisotropy (PMA) [40]. Such modifications of materials properties through ionic motion is highly effective in tailoring interfacial characteristics and consequently physical and chemical properties [41-46]. Hydrogen based magneto-ionics is particularly appealing, comparing to mostly oxygen-based systems studied so far, due to the superior reversibility and speed [40]. Yet quantitative understanding of the mechanism is still lacking because of the buried interfaces. Furthermore, besides PMA, other hydrogen-induced magneto-ionic functionalities remain largely unexplored. For example, it would be highly attractive to reversibly control the interfacial DMI, and in turn magnetic chirality, via hydrogen transport, especially given the high mobility of hydrogen in solids. Therefore, there is a critical need to confirm and quantify the DMI induced at hydrogen/ferromagnet interfaces and explore its effect on spin textures.

In this article, we report direct observation of a hydrogen-induced DMI at the surface of ferromagnets at room temperature and a reversible control of the magnetic chirality and spin textures in the ferromagnets. The interfacial DMI induced by chemisorbed hydrogen is quantitatively examined in perpendicularly magnetized Ni/Co/Pd/W(110) multilayers where the sign and magnitude of the DMI can be tuned by controlling the Pd layer thickness. When the multilayer has very weak Pd-like DMI (left-handed chirality), we discover a chirality transition of magnetic domain walls in the Ni/Co layers from left-handed to right-handed upon hydrogen chemisorption, indicating that the chemisorbed hydrogen on top of the surfaces introduces finite DMI which favors right-handed chirality, which is supported by density functional theory (DFT) calculations. We further demonstrate that the magnetic chirality of domain walls can be sensitively and reversibly switched between right-handed/left-handed during the chemisorption/desorption of hydrogen. Our results extend the picture of the interfacial DMI to the lightest element and enriches the hydrogen-related design of chiral spintronics and magneto-ionic devices.

## II. RESULTS AND DISCUSSION

### A. Reversible chemisorption/desorption of hydrogen on Ni(111) and Co(0001) surfaces

The essential hydrogen/metal interface is realized via chemisorption of hydrogen on solid surfaces of Ni(111) and Co(0001). In the dissociative adsorption of molecular hydrogen as used in this study (see Appendix A), prior experimental work showed that the hydrogen atoms adsorb favorably on the top surface (as opposed to diffusion into subsurface binding sites) due to the presence of a chemisorption energy well [47,48], which is also corroborated in more recent analyses using DFT for the case of Ni [49] and Co [50]. Experimental results show that in the case of Ni(111), the hydrogen atoms occupy three-fold hollow sites with a Ni-H bond length of $(1.84 \pm 0.06)$ Å, corresponding to an overlayer-substrate spacing of $(1.15 \pm 0.10)$ Å [51], and DFT results find that the binding geometry for hydrogen on close packed Co is very similar [50].

In our experiments the hydrogen coverage is monitored by measuring the chemisorption-induced work function shift $\Delta\varphi$ using Low-Energy Electron Microscopy (LEEM) [48]. The LEEM is a powerful tool to measure the work function of material surfaces by fitting LEEM IV curves (Fig. 1a) [52] (see Appendix B). We observed a work function increase of $\Delta\varphi \approx 120$ meV on a (111) oriented Ni film upon 0.9 Langmuir hydrogen exposure (180 seconds at $5 \times 10^{-9}$ torr ) at room temperature (Fig. 1a) (see Appendix A). This significant work function shift is in excellent agreement with prior work [53], where a shift of $\Delta\varphi \approx 139$ meV was reported to occur upon hydrogen adsorption on a Ni(111) surface at $41\,°C$ with hydrogen pressure set to $5 \times 10^{-9}$ torr, and with DFT calculations (Appendix D) where a shift of $\Delta\varphi \approx 141$ meV is found for a full atomic monolayer (ML) H coverage.

To explore the possible reversibility at room temperature, the evolution of $\Delta\varphi$ is monitored during cycles of turning hydrogen partial pressure ON and OFF, where the ON state refers to $5 \times 10^{-9}$ torr of hydrogen and OFF refers to the base pressure (below $4 \times 10^{-11}$ torr). For the hydrogen covered Ni(111) surface, prior flash desorption work identified two desorption maxima around 290-310 K ($\beta_1$ state) and 370-380 K ($\beta_2$ state) [53,54], where the $\beta_2$ feature is only observed at low hydrogen coverage and saturates at 0.5 ML hydrogen, and the $\beta_1$ feature is observed at higher coverage and saturates at 1 ML hydrogen [54]. Here the depth of the chemisorption well is hydrogen-coverage dependent due to the repulsive planar H-H interaction [48,54], which tends to lower the hydrogen binding energy at higher hydrogen coverage [48], resulting in different depth of the chemisorption well as $\Delta E_{\beta_1} = 0.43$ eV and $\Delta E_{\beta_2} = 0.50$ eV [47]. Because the desorption temperature of the $\beta_1$ state is just above room temperature, spontaneous hydrogen desorption at room temperature is expected in ultrahigh vacuum (UHV) condition [53]. Figure 1b shows the work function shift $\Delta\varphi$ on a Ni(111) surface as a function of time over four ON(3min)/OFF(10min) cycles. The plot shows the gradual work function increase of $\Delta\varphi \approx 120$ meV during the first hydrogen exposure (0.9 Langmuir), and reversible oscillations of $\Delta\varphi$ during the subsequent ON/OFF cycles with an amplitude of about $\pm40$ meV. The known dependence of $\Delta\varphi$ on the hydrogen coverage [48,53] indicates that chemisorption of hydrogen on Ni(111) is indeed partly reversible at room temperature, and desorption is likely limited to the $\beta_1$ state (high hydrogen coverage sites) [53,54]. Consistent with prior literature [53], our result indicates that roughly one third of hydrogen can be reversibly chemisorbed/desorbed on a Ni(111) film surface at room temperature and under ultrahigh vacuum conditions. Note that this coverage ratio may vary as a function of hydrogen dose and pressure [53].

Hydrogen chemisorption also occurs on the Co(0001) surface, where temperature programmed thermal desorption measurements indicated desorption maxima with coverage dependent positions around 325-370 K ($\beta_1$ state) and 400-420 K ($\beta_2$ state) [55], somewhat resembling the case of Ni(111). Similar to Ni(111), we found that cyclical hydrogen chemisorption/desorption on a Co(111) film is associated with a reversible work function change, albeit the amplitude is smaller with $\Delta\varphi \approx \pm20$ meV. Figure 1c plots time-dependent $\Delta\varphi$ measurements over four ON(3min at $5 \times 10^{-9}$ torr )/OFF(10min) cycles. The observed spontaneous hydrogen desorption from Co(0001) films at room temperature is consistent with the detailed thermal desorption study of this system reported in Ref. [55].

For DMI measurements described in detail below, we will use Ni/Co/Pd/W(110) multilayer samples. Here we first discuss the hydrogen chemisorption properties of such structures. Interestingly, we find that the hydrogen coverage ratio that results in cyclical chemisorption/desorption at room temperature can be greatly enhanced on these multilayer structures, compared to the single-element films described above. Figure 1d shows the evolution of $\Delta\varphi$ on the surface of a Ni(1)/Co(3)/Pd(2)/W(110) multilayer, where the numbers in brackets stand for layer thickness in ML, over an identical hydrogen ON/OFF cycle as shown in panels b and c. We find that the initial work function rise of $\Delta\varphi \approx 125$ meV upon hydrogen exposure (3min at $5 \times 10^{-9}$ torr) is comparable to $\Delta\varphi$ observed on Ni(111) (~120 meV). However, the amplitude of work function oscillations during the subsequent hydrogen pressure cycles is around 80 meV, about two thirds of the initial $\Delta\varphi$. This amplitude is almost twice that observed in the thicker (6 ML) Ni(111) film (Fig. 1b). The element Pd is known for its large bulk hydrogen adsorption capacity and one might surmise that the presence of 2ML Pd underneath the Ni/Co bilayer has something to do with the observed enhancement of hydrogen induced work function change. However, using a Ni(1)/Co(3)/Pd(20)/W(110) sample with a ten-fold thicker Pd layer, we observe that the $\Delta\varphi$ evolution induced by identical hydrogen ON/OFF cycles is almost identical as in the sample with just 2ML Pd (Fig. S1 [56]). This suggests that the large $\Delta\varphi$ ON/OFF ratio originates from the top Ni/Co bilayer, and not from the Pd layer. Moreover, the enhanced $\Delta\varphi$ ON/OFF ratio at room temperature, compared to the single element films, is likely related to the different hydrogen binding energy on the bilayer composed of closed-packed Co surface with a 1ML Ni overlayer [57]. Exactly how multilayer structure affects hydrogen desorption at various coverages may merit further investigations using temperature programmed thermal desorption, which exceeds the scope of this paper. An even greater $\Delta\varphi$ ON/OFF ratio can be achieved on the same Ni(1)/Co(3)/Pd(2)/W(110) structure at elevated temperatures. Figure 1e shows that when the sample is held at ~90 °C then in the hydrogen OFF part of the cycles the work function nearly fully recovers to the initial value of the hydrogen-free surface. As a result, the ratio of hydrogen coverage extrema in the ON/OFF cycles is on the order of ~90% of the initial work function rise. This indicates that the sample temperature of 90 °C is sufficient to activate the hydrogen desorption process related to the 2$^{nd}$ desorption maximum ($\beta_2$ state, the higher temperature desorption peak in the low hydrogen coverage case), which is comparable to the reported $\beta_2$ state at 380 K in the hydrogen/Ni(111) system [53]. Note that our observed initial work function rise of $\Delta\varphi \approx 50$ meV at 90°C is also in reasonable agreement with the value of $\Delta\varphi \approx 40$ meV reported in Ref. [53] for Ni(111) at 89°C in $5 \times 10^{-9}$ torr hydrogen.

**B. Exploring interfacial DMI induced by chemisorbed hydrogen**

Direct measurement of magnetic chirality is one of the major approaches to unravelling the interfacial DMI [12]. For instance, ground-breaking observations of cycloidal spin spirals using spin-polarized scanning tunneling microscopy have revealed the role of the interfacial DMI on magnetic chirality as well as the period of the spin spirals [5,6,9,10,13]. More recently, observation of magnetic chirality in magnetic domain walls also allows the quantification of the magnitude and sign of the interfacial DMI [16,25,58]. A particularly versatile approach to measure the DMI at the top interfaces of magnetic multilayers emerges when the magnitude and sign of the effective DMI induced at buried interfaces within the structure can be tuned predictably and accurately. This can be done by using hybrid substrates composed of a bulk crystal coated with a spacer layer where the crystal and spacer induce a DMI of opposite sign, such as Ir/Pt(111) [58], or Pd/W(110) [27]. The advantage of using a tunable-DMI substrate in this fashion was previously demonstrated in quantifying the DMI induced by chemisorbed oxygen on the Ni(111) surface [27]. Here, we test the DMI induced by chemisorbed hydrogen on the top surface of Ni(1)/Co(3)/Pd($t_{Pd}$)/W(110), where the effective DMI in the buried interfaces favors left-handed Néel chirality (Pd-like) at thick Pd thickness $t_{\text{Pd}}$, and right-handed Néel chirality (W-like) [14] at thin $t_{\text{Pd}}$.

What makes this method advantageous for quantifying even rather weak DMI contributions is the fact that the magnitude and sign of the effective DMI of the buried interfaces can be fine-tuned right around the point of null-DMI. Here we tracked the magnetic chirality evolution upon hydrogen chemisorption on various samples with different initial chirality. We observe a clear hydrogen-induced chirality switching in samples with Pd spacer layer thickness $t_{\text{Pd}}\sim2.09\text{ML}$, where the effective DMI of the hydrogen-free multilayer is weakly Pd-like (left-handed). Figure 2a shows a SPLEEM image of the sample in the as-grown state, where the domain wall magnetization preferentially points from gray domain ($-\mathbf{M}_z$) to the black domain ($+\mathbf{M}_z$), corresponding to left-handed Néel chirality. Upon hydrogen chemisorption, Figure 2b shows that the same domain wall evolves to right-handed Néel chirality (now the domain wall magnetization predominantly points from black domain ($+\mathbf{M}_z$) to gray domain ($-\mathbf{M}_z$) in Fig. 2b). We denote this switching of the magnetic chirality as the chirality transition. For a more quantitative analysis, we measure domain wall chirality in a statistically significant number of image pixels along the domain wall center-line. Defining the parameter $\alpha$ as the angle between the domain wall normal direction $\mathbf{n}$ and the magnetization vector $\mathbf{m}$ at each point along the domain wall center-line (see inset in panel c), histograms of this angle $\alpha$ measured from SPLEEM images represent the statistics of domain wall chirality [25,58]. Figure 2c/2d show that, before/after a 0.9 Langmuir hydrogen exposure, the peak at $\alpha\sim0°$ in panel c indicates left-handed Néel chirality, whereas the peak at $\alpha\sim180°$ in panel d indicates right-handed Néel structure. This statistical approach allows quantification of the chirality transition as shown in Figure 2e, where the average domain wall chirality before and after 0.9 Langmuir hydrogen exposure is plotted for several samples, as a function of Pd spacer layer thickness $t_{\text{Pd}}$. Note the hydrogen coverage resulting from this dose at room temperature can be roughly estimated as $t_{\text{H}}=(0.6\pm0.1)\text{ ML}$ with respect to the planar atomic density of Ni(111) (see Appendix A).

When the Pd spacer layer is too thin and the effective DMI remains W-like (right-handed), as in the $t_{\text{Pd}}=2.00$ ML and $t_{\text{Pd}}=2.05$ ML measurements, then the domain wall chirality remains completely unaffected by hydrogen chemisorption because the induced DMI at the hydrogen/Ni interface and the effective DMI in Ni/Co/thin-Pd/W have the same sign (both right-handed). Likewise, when the Pd spacer layer is too thick, as in the $t_{\text{Pd}}=2.15$ ML sample, then the Pd-like effective DMI (left-handed) is sufficiently strong to dominate the domain wall spin texture, and the chirality of the wall remains unaffected even after hydrogen chemisorption because the sign of the effective DMI (left-handed) will not change with the additional weak hydrogen-induced right-handed DMI. However, when the initial DMI is sufficiently weak, as in the samples with $t_{\text{Pd}}$ = 2.08 ML, 2.09 ML and 2.10 ML, then hydrogen chemisorption induces a transition of the domain wall chirality, clearly revealing the right-handed DMI induced at the hydrogen/Ni(111) top interface. Here the magnitude of the effective DMI is considered to change roughly linearly as a function of sub-monolayer Pd thickness variation [27], because the typical length scale of the domain walls in our experiment is much larger than the size of third-monolayer Pd islands on top of the completed second-layer Pd film (see Appendix A). It is interesting to consider whether atomic-scale roughness, such as atomic steps surrounding the islands that make up incomplete atomic monolayers might introduce appreciable DMI. However, if one particular kink-site at the edge of a monolayer island induces a certain amount of DMI, then another mirror-symmetric kink site on the opposite side of the island would induce equal and opposite amount of DMI because of the mirror-symmetric atom arrangement. In absence of a symmetry-breaking condition, any arrangement of atoms is energetically degenerate with its mirror arrangement. For this reason populations of left-handed and right-handed DMI-inducing step-segments are expected to occur with equal frequencies and must cancel each other. That said, in principle crystals can be cut in planes that expose chiral kinks such as vicinal surfaces [59-61]. However, the epitaxial layers grown on the W(110) surface studied here are achiral and for these symmetry reasons atomic-scale roughness induced contributions to the effective DMI would vanish.

Note that the typical Néel- to Bloch-wall transition near zero DMI is suppressed because a weak in-plane uniaxial magnetic anisotropy in this system prevents Bloch-like alignment of domain wall magnetization along the W[1-10] direction [27]. The Néel components of the wall magnetization, however, are clearly sensitive to the sign of the DMI [62]. These results show that chemisorbed hydrogen on top of the Ni(111) surface introduces finite DMI favoring right-handed spin structures, i.e. the same sign of the DMI induced by overlayer Pt, Pd or oxygen [27] (left-handed is favored when ferromagnetic layer is on top).

## C. Estimation of the strength of chemisorbed hydrogen induced DMI

The systematic $t_{Pd}$ spacer layer thickness-dependent chirality studies summarized in Figure 2e allow us to estimate the magnitude of the hydrogen-induced DMI. The chirality evolution towards right-handedness is observed between 2.08 ML and 2.10 ML during 0.6 ML hydrogen chemisorption. Above 2.10 ML Pd, no significant chirality change can be observed as the initial effective Pd-like DMI now dominates and hydrogen induced DMI at the Ni(111) surface can no longer affect the chirality. This approach provides an opportunity to quantify the hydrogen induced DMI by linking it to the dependence of the initial DMI on the Pd spacer layer thickness $t_{Pd}$. Without hydrogen the achiral state of domain walls, where the effective DMI is essentially zero, occurs at $t_{Pd} \approx 2.08$ ML. Upon chemisorption of 0.6 ML hydrogen the achiral state shifts to $t_{Pd} = (2.095 \pm 0.004)$ ML . The relative change of the DMI in the Ni/Co/Pd/W(110) system as a function of the Pd layer thickness $t_{Pd}$ was previously quantified as $(0.41 \pm 0.17)$ meV/ surface atom  per monolayer $\Delta t_{Pd}$=1 ML [27]. For the DMI values estimated by SPLEEM-images-based evolution of the thickness-dependent domain wall [27,58], we note that the unit of the DMI vectors is given in meV per surface (interface) atom. The measurements summarized in Figure 2e show that the change of effective DMI induced by 0.6 ML hydrogen chemisorption on top of the Ni/Co/Pd/W(110) multilayer is equivalent to the change of the DMI induced by increasing the Pd spacer layer thickness by $t_{Pd} = (0.015 \pm 0.004)$ ML  in the absence of hydrogen. Therefore, the DMI induced by the chemisorbed hydrogen on Ni/Co/Pd/W can be estimated as:

$$(0.41 \pm 0.17) \times \frac{0.015 \pm 0.004}{0.6 \pm 0.1} \text{ meV/surface atom} = (0.01 \pm 0.005) \text{ meV/surface atom  for 1 ML}$$

equivalent hydrogen coverage. Note that only the chemisorption of hydrogen on the Ni surface and the role of the hydrogen/Ni interface on the DMI are considered here because the combination of chemisorption well and energy barrier of hydrogen bulk diffusion favors the occupation of chemisorption sites [54] in our approach of introducing low dose (0.9 Langmuir) hydrogen molecules when the sample is held at room temperature.

Figure 2f shows a comparison of the DMI induced by various elements adjacent to Ni. For instance, the chemisorbed hydrogen induced DMI is much weaker than the chemisorbed oxygen induced DMI on Ni, which is $(0.63 \pm 0.26)$ meV/surface atom at 1 ML equivalent oxygen coverage. The strength of the hydrogen-induced DMI is one to two orders of magnitude smaller than the DMI induced at Ni/transition metal interfaces, for example, $D_{Ni/Cu + Fe/Ni} = (0.15 \pm 0.02)$ meV/surface atom [16], $D_{Ni/W} \approx 0.24$ meV/surface atom [62], $D_{Ni/Ir} = (0.12 \pm 0.04)$ meV/surface atom [58], $D_{Ni/Pt} = (1.05 \pm 0.18)$ meV/surface atom [58]. The hydrogen-induced DMI is also much weaker than the DMI induced at the Co/graphene interface, which is $(0.16 \pm 0.05)$ meV/surface atom [25]. Note that here we only compare DMI measured in SPLEEM-based experiments using methods described in Refs. [16], [62] and [27], to avoid possible systematic measurement biases resulting from the use of different methods [12]. The element-dependent magnitude of the DMI is related to different mechanisms [12]. In the Fert-Levy model [28,63], the DMI scales with the strength of spin-orbit coupling of the adjacent heavy-metal [28], and with the degree of orbital hybridization at the 3$d$-5$d$ interface as well as 3$d$ band lineup dictated by the Hund's first rule [30]. The graphene-induced DMI is dominated by the Rashba effect and its magnitude scales with the Rashba coefficient [25]. The oxygen-induced DMI is related to the charge transfer and hybridization at the interface [27,31], and the plausible mechanism of hydrogen-induced DMI is discussed in section D below.

Figure 2g summarizes DFT analysis of the H-induced change of the micromagnetic DMI-strength $D$ for the system Ni(1)/Co(3)/Pd($d_{Pd}$)/W (for details of the calculations see Appendix D). Since the DMI depends on structural details the total DMI is calculated as the sum of layer-decomposed contributions. The DMI of Ni and Co (Pd and W) layers are calculated in the 250 (275) pm lattice constant model. We find that in the absence of Pd, the contribution of W to the DMI is negative, favoring a right-handed domain-wall, consistent with the analysis of Fe on W(110) [14]. The DMI of Pd favors left-handedness (like Pt), and with increasing Pd thickness, $D$ becomes larger and changes sign in Pd/W. The DMI of Ni and Co favors left-handedness (like Pd), but the magnitude is smaller than that of Pd and W, consistent with the lower atomic number Z and the subsequently smaller spin-orbit interaction. Adding up all contributions, we obtain the total DMI of Ni(1)/Co(3)/Pd($d_{Pd}$)/W with a sign change at a Pd thickness of $d_{Pd} \approx 1$ ML. In comparison to the Fig 1H of Ref. [27], the Pd thickness of achirality (thickness of zero DMI) moved by about 0.5 ML due to our refined structural model treating the Pd and W layers in the 275 pm lattice constant instead of 250 pm. Adding 1ML H on Ni(111) we find that the total strength $D$ is reduced and H acts like an additional contribution favoring the right-rotating domain wall, thus requiring a larger Pd thickness to reach the achiral point. All these are consistent with the experimental facts, but in comparison with the experimental data, the theoretical DMI values are somewhat larger. The experimental thickness to achirality is about $d_{Pd} \approx 2$ ML, and theoretically the hydrogen-induced shift equivalent is found to be 0.35 ML of Pd, while in the experiment it is about 0.015 ML of Pd. We attribute these quantitative discrepancies to the difficulty in accurately modeling the true structure of the materials stack.

### D. Hydrogen-assisted reversible control of the chirality

The observation of substantial reversibility of hydrogen chemisorption by desorption in clean UHV at room temperature, together with the observed hydrogen-induced switching of domain wall chirality, suggests the possibility to reversibly switch the domain wall chirality by hydrogen chemisorption/desorption cycles. To test this possibility, we use SPLEEM to continuously monitor the domain wall magnetization in a Ni(1ML)/Co(3ML)/Pd(2.09ML)/W(110) multilayer, while periodically cycling between $5 \times 10^{-9}$ torr hydrogen pressure for 3 minutes and negligible hydrogen pressure (UHV base pressure) for 10 minutes. Figure 3a shows the evolution of the domain wall chirality in the four cycles, where the chirality switched from predominantly left-handed to predominantly right-handed upon the 1st hydrogen chemisorption (see the definition of the chirality in Fig. 3a), and the chirality partially evolves toward left-handedness/right-handedness during "H OFF"/"H ON" states for the rest of the cycles. Figure 3b shows the statistics of this domain wall switching experiment, tracking reversibility of the chirality over four cycles at room temperature (see Appendix C). This magnetic chirality measurement is correlated with a hydrogen coverage measurement, as monitored by tracking the work function change of +120 meV for the 1st H-on state and $\pm 80$ meV for the subsequent cycles. These results indicate that the hydrogen coverage changes shown previously in Fig. 1d indeed reversibly affect the DMI of the system so as to switch the sign of the effective DMI as well as the domain wall chirality. In this experiment, the chirality reversal during the H-OFF state is imperfect in the sense that a small fraction of domain wall sections remains in the right-handed state corresponding to the hydrogen-induced DMI. It is plausible that these minor imperfections in the chirality switching are due to a combination of defect-induced pinning and the weaker DMI associated with residual hydrogen coverage due to incomplete desorption in the 10-minute OFF cycles.

In the end, the physical origin of the chemisorbed hydrogen-induced change in the DMI is the change in the spatial asymmetry of the wave function in the transition metals, especially in Ni surface atoms. The hydrogen causes this change by a slight outward relaxation of the nickel position above the Co film (see Appendix D), induced by a small charge transfer from nickel to hydrogen and by the polarization and rehybridization of the orbitals in nickel. The electrical surface dipole moment and the work function are the corresponding physical quantities that provide information about these changes. In fact, a linear relationship between the work function change and the DMI was demonstrated for Pt/Co/transition metals trilayers [64], and a linear relationship between the electronegativity, the electric dipole moment and the DMI was proposed in Ref. [29]. To shed more light on these relationships, we compare the changes in DMI (H: 0.01 $\pm$ 0.005 meV/1ML H, O: 0.63 $\pm$ 0.26 meV/1ML O), work-function (H: ~0.12 eV, O: ~0.70 eV) and Pauling electronegativity differences (H: 0.29, O: 1.53, relative to Ni) between the hydrogen (H:) and oxygen (O:) [27] terminated NiCoPdW systems (the electronegativities of H, O, and Ni are 2.20, 3.44, 1.91, respectively). Note that the electronegativity of both hydrogen and oxygen is larger than that of nickel, which is consistent with the measurements that the work function on the surface of NiCoPdW increases upon adsorption of both hydrogen and oxygen. Also, the electronegativity of oxygen is larger than that of hydrogen, and the strength of the right-handed DMI increases, more significantly for the oxygen system than for the hydrogen system. Indeed, the ratios of the work function shift ( $R_{\Delta\varphi} = 0.70/0.12 \approx 5.8$ ) and the changes of the electronegativities ( $R_{\Delta\chi} = 1.53/0.29 \approx 5.3$ ) between the oxygen system and the hydrogen system are about the same, i.e. $R_{\Delta\varphi}/R_{\Delta\chi} \approx 1$. Evidently, for these systems the change of the electronegativity is an excellent measure for the estimation of the change of the work function. Considering the ratios between the change in DMI to the change in work function, one obtains 0.63 meV / 0.70 eV for the oxygen system, i.e., a change in DMI of ~1 meV per 1 eV change in work function. For the hydrogen system, one gets a value which is more than 10 times smaller (0.01 meV/0.12 eV). Also, the same ratios are obtained if the work function changes are replaced by the electronegativities. Testing whether one-to-one relationships between DMI and work function change [64] or electronegativity [29] hold more generally goes beyond the scope of this paper. We speculate that either the changes induced by the oxygen are so strong that we are no longer in the linear regime or the DMI value of the hydrogen system is so small due to the compensation of the different contributions.

Chemisorption of hydrogen occurs on many transition metals, in particular a considerable hydrogen-induced dipolar moment appears (via the observation of a work function shift) on the surfaces of ferromagnetic metals such as cobalt, nickel and iron or 4$d$/5$d$ metals, and we expect that chemisorbed hydrogen induced DMI can be generally observed on ferromagnetic thin films. However, the reversibility demonstrated in Figs. 1 and 3 may require additional testing for each specific case. Furthermore, we note that hydrogen is the lightest element of the periodic table. It is tiny, has no chemical aggressiveness, and is not as electronegative as oxygen. In this sense hydrogen is rather gentle, might be the ideal chemical element that enables the best control of the DMI needed to change the spatial properties of spin textures such as skyrmions, much more precisely than oxygen.

The sensitive and reversible switching of the DMI and chiral spin texture via hydrogen chemisorption is highly relevant for chiral spintronics, such as racetrack memories based on domain walls or skyrmions, where hydrogen-induced DMI may be used to manipulate the chiral domain wall motion by controlling the chirality, or to sensitively control the skyrmion size over orders of magnitude [22], a feature particularly interesting for neuromorphic computing. One key advantage is that the switching via chemisorption may be done in a tunable and contactless fashion, without requiring electrical leads being attached to the device. This is particularly attractive for complex device geometries such as 3-dimensional racetrack memory [65] and networks [66] involving numerous domain walls or skyrmions. Note that the role of hydrogen on the DMI is not limited to surfaces as demonstrated in this paper, and rich possibilities exist where hydrogen may occupy sites inside the bulk, such as hydrogenated Fe/Ir(111), which may also alter the DMI via lattice spacing changes [36]. These hydrogen-based results are also relevant to the emerging field of magneto-ionics, which has so far been largely based on oxygen ions and vacancies. They not only significantly expand on the magnetic functionalities that can be controlled magneto-ionically, but also offer exciting potentials for completely reversible and energy-efficient switching. We speculate that the hydrogenation of interfaces of films consisting of magnetic and high-spin-orbit materials opens a vista to new magneto-ionic or memristive functionalities. For example, electrical control of hydrogen coverage through magneto-ionic means could lead to a change of DMI and consequently spin textures, such as chiral domain wall motion or skyrmion size; in the former case the chiral domain wall motion may be useful for magnetic logic devices as well as artificial synapses [67], and in the latter case magnetoresistive properties of skymion-based devices may be relevant for memristors. Importantly, using the chemisorption mechanism, the hydrogen ions can be driven across atomic distances to contact the relevant ferromagnet surface in solid state devices (and reduce to atomic form [40]), but without actually penetrating the surface and causing any irreversible changes, thus leading to excellent reversibility, endurance and potential for high speed.

## III. CONCLUSION

In summary, we report direct and quantitative observation of a hydrogen chemisorption induced DMI on ferromagnet surfaces at room temperature, which can be used to sensitively and reversibly switch chiral domain walls. We find that the chemisorption/desorption ratio of the hydrogen is greatly enhanced by combining 1ML Ni and 3ML Co at the top of the multilayers and, even under constant room temperature conditions, the reversibility of hydrogen chemisorption can reach as much as two thirds of the initial hydrogen coverage. We observe that the hydrogen chemisorption induces a finely controllable, reversible and non-volatile chirality transition of magnetic domain walls in the Ni/Co/Pd/W(110) system at room temperature. This chirality control is attributed to the hydrogen induced DMI, which is experimentally quantified. These results help to advance the understanding of the DMI induced by light elements, and open up new device potentials in chiral spintronics and magneto-ionics.

## ACKNOWLEDGMENTS

We thank Hongying Jia and Tianping Ma for insightful discussions, and Steven E Zeltmann for help with the TVD implementation. This work has been supported in part by the NSF (DMR-1905468 and DMR-2005108), the UC Office of the President Multicampus Research Programs and Initiatives (MRP-17-454963), and SMART (2018-NE-2861), one of seven centers of nCORE, a Semiconductor Research Corporation program, sponsored by the National Institute of Standards and Technology (NIST). Work at the Molecular Foundry was supported by the Office of Science, Office of Basic Energy Sciences, of the US Department of Energy under contract no. DE-AC02-05CH11231. Work at NJU has been supported by the National Key R&D Program of China (Grant No. 2017YFA0303202), the National Natural Science Foundation of China (Grants No. 11734006 and No. 11974165). R.L.C., A.K.S. and R.W. acknowledge financial support by the European Union via an International Marie Skłodowska-Curie Fellowship (grant No. 748006 - SKDWONTRACK). C.O. acknowledges support from the US Department of Energy Early Career Research Program. S.B. acknowledges financial support from the DARPA TEE program through grant MIPR (# HR0011831554) from DOI, from Deutsche Forschungsgemeinschaft (DFG) through SPP 2137 "Skyrmionics" (Project BL 444/16), the Collaborative Research Centers SFB 1238 (Project C01).

A.L.F.C. and A.K.S. thank the Advanced Research Projects Agency-Energy (ARPA-E), U.S. Department of Energy, under Award Number DE-AR0000664.

**APPENDIX A: Sample preparation and hydrogen exposure**

The SPLEEM experiments were performed at the National Center for Electron Microscopy of the Lawrence Berkeley National Laboratory. All samples were grown in the SPLEEM chamber under ultra-high vacuum (UHV) conditions, with a base pressure better than $4.0 \times 10^{-11}$ torr. The W(110) substrate was cleaned by cycles of flashing to 1,950°C in $3.0 \times 10^{-8}$ torr $O_2$, followed by a final flashing at the same temperature to remove oxygen. Ni, Co and Pd layers were deposited by physical vapor deposition from electron beam evaporators when the substrate is held at room temperature, and the film thicknesses of Ni, Co and Pd layers were calibrated by monitoring the oscillations of the LEEM image intensity associated with atomic layer-by-layer growth [27]. In contrast to the significantly larger Pd island structures grown at high temperature [68], in our room-temperature sample preparations the strong layer-by-layer electron reflectivity oscillations together with the observation of featureless LEEM images indicate that the typical size of next-monolayer Pd islands at the growth front of the Pd films is smaller than the spatial resolution at the magnification used in these experiments (image pixel size is ~22 nm at 10 μm field of view used in this work). Note that the Pd thickness with zero effective DMI ($t_{\mathrm{Pd}} \approx 2.08$ ML) here is slightly thinner than that reported in ref. [27] ($t_{\mathrm{Pd}} \approx 2.46$ ML), which is possibly due to slight differences in experimental conditions.

Hydrogen exposures were realized by leaking of high-purity hydrogen (99.999%) at a pressure of at $5 \times 10^{-9}$ torr. The pressure of hydrogen reading of the ionization gauge has been corrected by a factor of 0.46. No noticeable change was observed in the LEED pattern upon hydrogen chemisorption at room temperature, which is consistent with ref. [51]. On Ni(111), the maximum work function shift occurs at the hydrogen coverage of 0.5-0.6 ML [51], and volumetric measurements reveals that the saturation coverage of chemisorbed hydrogen on Ni(111) is ~0.7 ML at room temperature [53]. Therefore, the hydrogen coverage on the surface of Ni/Co/Pd/W(110) is estimated based on the work function shift measurement with a maximum work function shift ($\Delta\varphi \approx 125$ meV, Fig. 1d), which roughly corresponds to 0.5-0.7 ML hydrogen overlayer.

**APPENDIX B: Time-dependent work function measurement**

The work function is determined by fitting the LEEM IV spectrum (image intensity vs incident energy of electrons, see Fig. 1a) with a complementary error function $erfc$ (Start voltage)[52]. The value where the drop-off occurs, $V_S^0$, represents a measurement of the sample work function given by $\phi_{\mathrm{sample}} = V_S^0 + E_C^0$, where $E_C^0$ represents the peak of the electron distribution emitted from our photocathode (p-type GaAs crystal activated with $CsO_x$). The emission of the GaAs cathode of SPLEEM is set to 100nA to optimize the energy spread to about 180 meV (full width at half maximum) and $E_C^0 \sim 1.4 - 1.5$ eV measuring a reference surface such as Highly Oriented Pyrolytic Graphite (HOPG). Time-dependent work function measurements were performed by recording the reflectivity of low energy electrons while sweeping the start voltage in a loop (Fig. S2 [56]). In order to record the work function changes during hydrogen adsorption/desorption at the surface, the start voltage was swept from 1.5 V below to about 2 V above the intensity drop-off using 50 mV voltage steps and an image integration time of 250 ms. Relative changes in the work function over time can be detected with very-high sensitivity down to about 5 mV given by the shift of the centroid of the gaussian distribution extracted by the $erfc$(Start voltage) fitting.

**APPENDIX C: Time-dependent in-plane domain wall analysis**

Due to the noise present in the individual in-plane domain wall images, we used standard image denoising methods to provide a more accurate estimate for the magnetization presented in Fig. 3. The measured images were denoised by 3D total variational denoising (3D-TVD), using a Matlab implementation and 3D extension to the methods given in ref. [69]. After normalizing the data to have a mean intensity of zero and a standard deviation of one, we used regularization parameters of $\mu$ = [2 2 1] and $\lambda$ = [1/8 1/8 1/16] for the dimensions of x,y and time respectively. FISTA acceleration was used to speed convergence. The regularization was applied isotropically to the x and y directions. After the TVD was applied, we normalized the images to have a mean of zero and the boundary contrast to have an approximate range of -1 to +1.

**APPENDIX D: DFT results and analysis**

All first-principles calculations of clean and hydrogen covered Ni(1)/Co(3)/Pd(n)/W(110) systems were based on DFT. All calculations, approximations to exchange and correlation functional, first-principles methods, computational procedures and computational parameters are consistent with the calculations in Ref. [27]. We have investigated systems varying the thickness of Ni between 0 and 3 atomic layers and of Pd between 0 and 3 layers keeping the number of Co layers fixed at 3 atomic layers.

*Structural Model*: Since the substrate is a (110) oriented bcc W crystal, Pd, Co and Ni grow as fcc (111), and the lattice constants of Ni and Co are considerably different from Pd, we expect that some adaptation of the atoms and some strain release may take place on the way from the W-surface to the Ni-layer, which is difficult to capture in full detail in our structural model. We have modeled all systems by pseudomorphically (111) stacked transition-metal layers with a stacking order of the preferred bulk ordering for each metal. The system was approximated by an asymmetric stack of layers where the W-substrate is modeled by 5 layers of W. We studied results for two different structural models (a) with the in-plane lattice constants $a_{\mathrm{IP}} = 250$ pm, and (b) with $a_{\mathrm{IP}} = 275$ pm, the former is related to the average of the experimental bulk lattice constant of Ni ($a_{\mathrm{Ni}} = 352$ pm and $a_{\mathrm{IP\_Ni}} = a_{\mathrm{Ni}}/\sqrt{2} = 249$ pm) and hcp-Co ($a_{\mathrm{IP\_Co}} = 251$ pm) and the latter to the average of the experimental bulk lattice constant of fcc-Pd ($a_{\mathrm{Pd}} = 389$ pm and $a_{\mathrm{IP\_Pd}} = 275$ pm) and the cubic bulk lattice constant of W ($a_{\mathrm{W}} = 316$ pm and $a_{\mathrm{IP\_W}} = a_{\mathrm{W}}\sqrt{3}/2 = 274$ pm). To most accurately model the W(110) substrate, we adjusted the W interlayer distances to obtain the correct experimental bulk volume. For simplicity, we considered a 100% coverage of H on the Ni surface. In the following, we relate all surface-related quantities to the results of the 250 pm model and all W and Pd related results to the 275 pm model.

*Results*: On the basis of structural optimization by energy minimization we found: (i) H adsorbs on the fcc hollow sites for the $a$=250 pm model. (ii) H reduces the bond strength between Ni and Co and increases the interlayer distance from 198 pm to 207 pm. This result is basically independent of the number of Pd layers. (iii) H reduces substantially the magnetic moment of Ni from $0.72\,\mu_B$ to $0.21\,\mu_B$ and by a tiny amount (2.2%) the magnetic moment of the first Co layer adjacent to Ni. (iv) The work function of W in the 275 pm model amounts to $\varphi_W = 5.40$ eV, in good agreement with the experimental work function of W(110), $\varphi_W = 5.22$ eV. (v) The work function of Ni/Co/Pd/W in the 250 pm model amounts to $\varphi_{Ni} = 5.28$ eV, in good agreement with the experimental work function of Ni(111), $\varphi_{Ni} = 5.35$ eV. The result is practically independent of the number of Pd layers. (vi) The work function change $\Delta\varphi$ upon adsorption of H on the surface of Ni for the system Ni(1)/Co(3)/Pd(2)/W is $\Delta\varphi \approx 141$ meV, which is in excellent agreement with the experimentally measured value $\Delta\varphi \approx 139$ meV for H on Ni(111) at 41 °C and $5 \times 10^{-9}$ torr hydrogen pressure [53] and within this work for which $\Delta\varphi \approx 125$ meV was found but for a smaller coverage of about 0.6 ML.

*DMI*: In determining the interface DMI and its changes upon hydrogen coverage, we proceeded analogously to the experimental procedure. We first calculated from DFT the DMI strength $D$ [61] of the micromagnetic DMI tensor, which was then expressed in the language of the atomistic spin-model, as in the experiment, with the energy expression $E_{DM} = \sum_{<ij>} \mathbf{D}_{ij} \, (\mathbf{S}_i \times \mathbf{S}_j)$ in terms of an effective DMI-vector $\mathbf{D}_{ij}$ between nearest neighbor (n.n.) pairs. Here *<ij>* denotes the summation over unique pairs of nearest-neighbor atoms. The effective atomistic nearest neighbor strength, $D_{n.n.}$, and the micromagnetic DMI strength, $D$, are related as $D_{n.n.} = 1/(3\pi a_{IP})\,D$, with $a_{IP} = 251$ pm (see the in-plane lattice constant $a_\parallel$ used to determine the dipolar energy [27]). Summing up all pairs $<ij>$ across the domain wall, $E_{DM}$ corresponds to the DMI energy in one domain wall and relates to the size of the effective nearest neighbor DMI vector $D_{n.n.}$ as $E_{DM} = -\sqrt{3}\pi D_{n.n.}$.

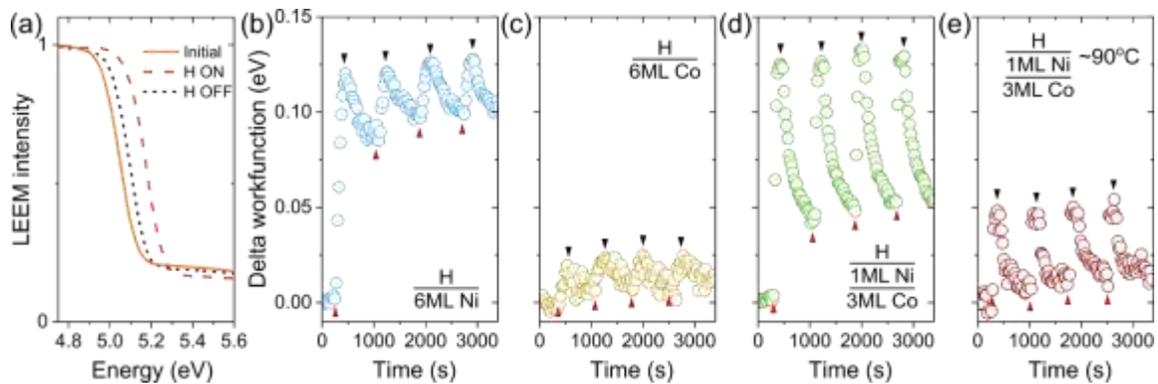

Figure 1. Room temperature observation of reversible chemisorption/desorption of atomic hydrogen on metal surfaces. (a), LEEM IV spectra on bare 1ML Ni/3ML Co (initial) and the same surface before/after the hydrogen exposure. Measuring the energy at which the reflectivity drops allows quantification of the work function. (b-d), Work function response on the surface of metals during the presence/absence of hydrogen at room temperature. Red/black triangles indicate the ON/OFF control of the hydrogen leak valve. Panel b: 6ML Ni, panel c: 6ML Co, panel d: 1ML Ni/3ML Co. (e), Work function response of 1ML Ni/3ML Co at ~90 °C, indicating ~90% chemisorption/desorption ratio.

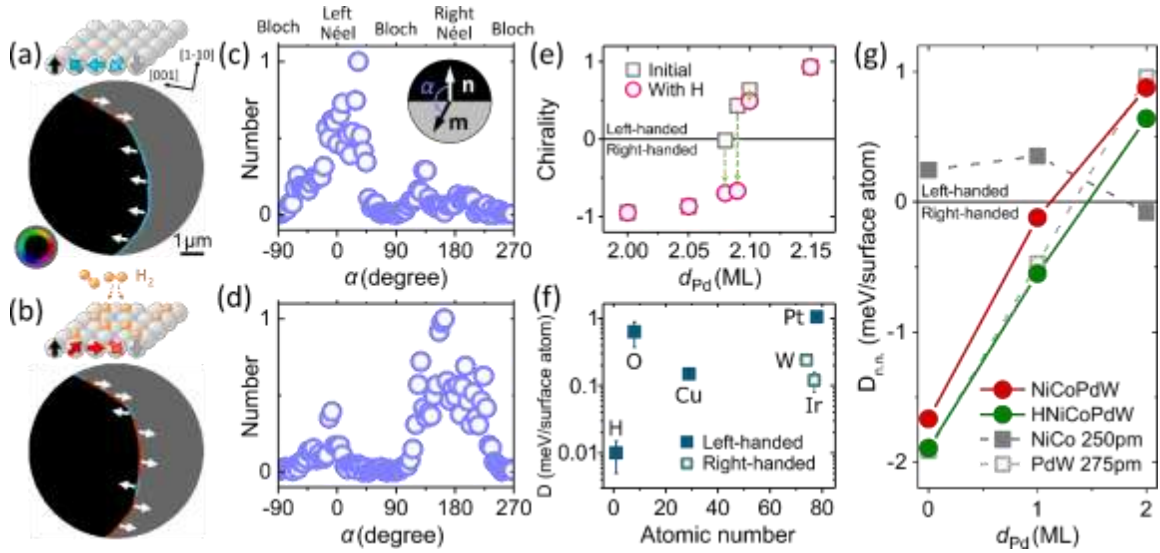

Figure 2. Exploring chemisorbed hydrogen induced Dzyaloshinskii–Moriya interaction. (a,b) Observation of hydrogen induced domain wall chirality switching in compound SPLEEM images of 1ML Ni/3ML Co/2.09 ML/W(110), panel a: as-grown (sketch shows left-handed walls in magnetic layers), panel b: with hydrogen exposure at $5 \times 10^{-9}$ torr (sketch shows right-handed walls upon hydrogen chemisorption). Black/gray area indicates perpendicularly magnetized up/down domains, colors indicate the in-plane orientation of magnetization in the domain wall region. (c,d) $\alpha$ Histogram of the SPLEEM image before (panel c)/after (panel d) the hydrogen exposure, $\alpha$ is the angle between domain wall magnetization **m** and domain wall normal vector **n** (insert). (e) Hydrogen exposure dependent evolution of Néel-type chirality at various Pd thicknesses. (f) Summarized values of DMI strength $D$ induced by various elements adjacent to Ni, all measured by the same SPLEEM-based method. (g) DFT calculation of Pd layer $d_{Pd}$ dependent DMI strength $D$ from various contributions: Ni/Co layers in the 250 pm structure model (solid square), Pd/W in the 275 pm model (hollow square), the sum of both contributions (red dots) and the total contribution including the chemisorbed hydrogen (H/Ni/Co in 250 pm model + Pd/W in 275 pm model) (green dots). Lines are guides to the eyes.

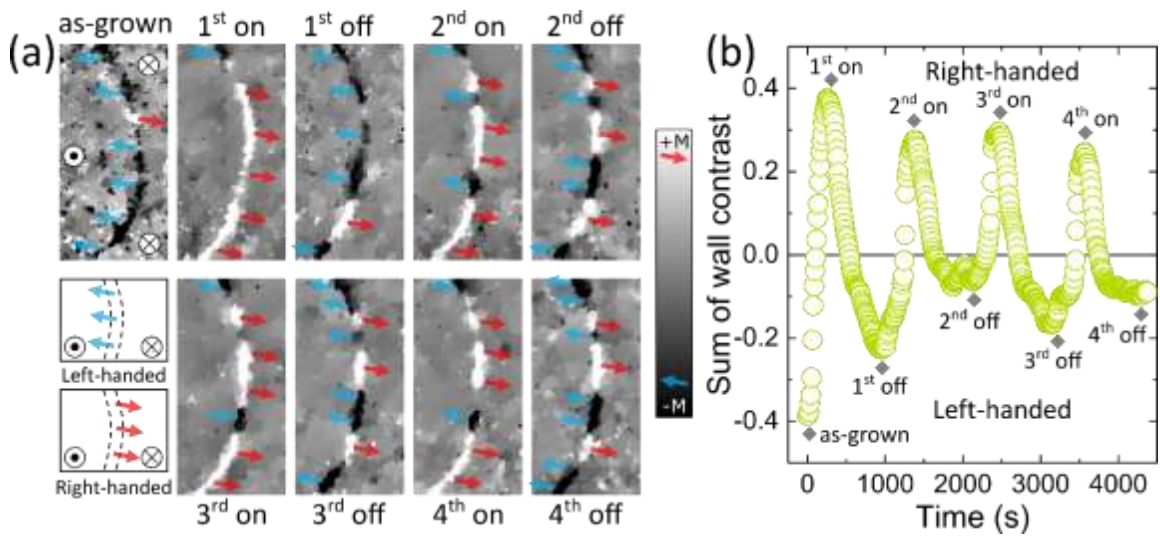

Figure 3. Reversible switching of magnetic chirality via hydrogen at room temperature. (a) time sequence of SPLEEM images of a domain wall in a Ni(1ML)/Co(3ML)/Pd(2.09ML)/W(110) system, hydrogen status is labelled above/below images. The in-plane magnetization in the domain wall region is rendered in gray-level according to the scale bar (right). Domains left and right of the domain wall are perpendicular magnetized. The magnetization in the left/right region points up/down, respectively. Magnetic chirality is highlighted by red/cyan arrows (see sketch). The field of view is $2\mu m \times 4\mu m$. (b) Evolution of average magnetic chirality (derived from the sum of the wall contrast). Gray diamonds indicate the timing of the images in a.

Supplemental Material for

Observation of hydrogen-induced Dzyaloshinskii–Moriya

interaction and reversible switching of magnetic chirality

Chen et al.

**Contents:**

**S1. Pd thickness dependent work function change**

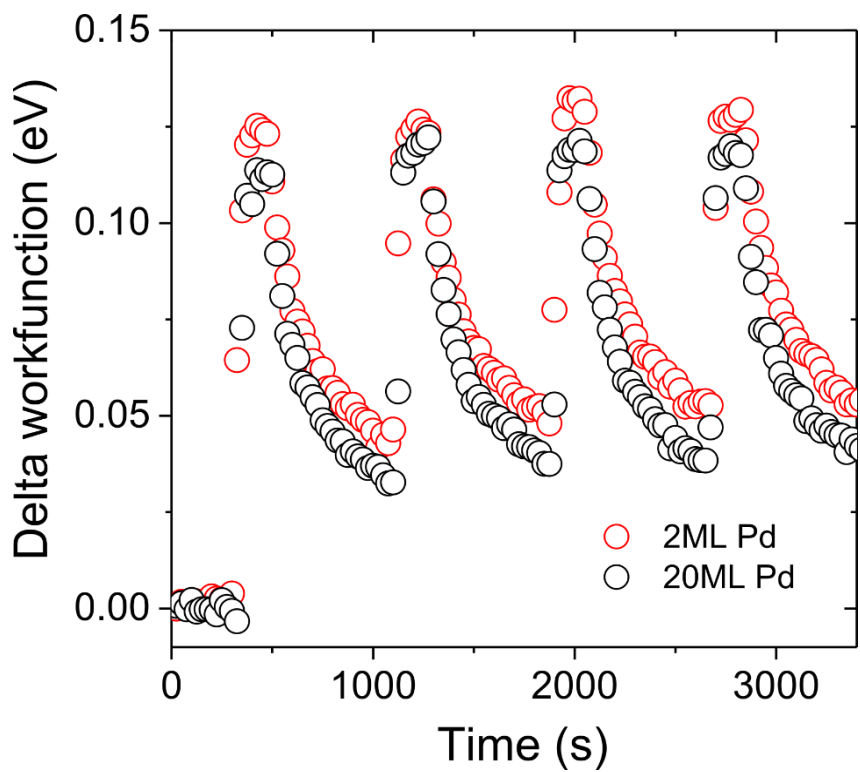

Figure S1. Work function response on the surface of metals during the presence/absence of hydrogen in Ni(1ML)/Co(3ML)/Pd(2ML)/W(110) and Ni(1ML)/Co(3ML)/Pd(20ML)/W(110) systems at room temperature.

**Fig. S2 Time-dependent work function measurement**

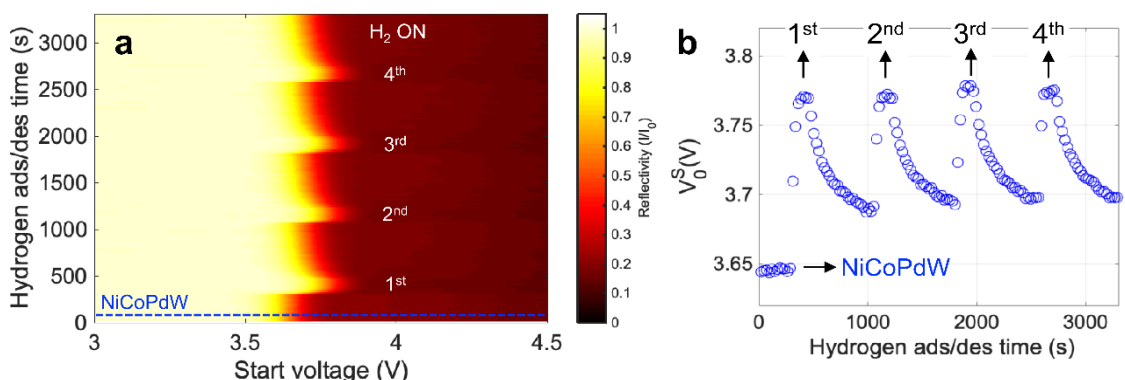

Figure S2. Evolution of the work function and electron reflectivity as a function of hydrogen adsorption/desorption for Ni(1ML)/Co(3ML)/Pd(2ML)/W(110). (a) The electron reflectivity as a function of electron energy (horizontal scale) and hydrogen adsorption/desorption time (vertical scale). The electron reflectivity across the full plot was normalized by the total 100% reflection given by the mirror electron mode (MEM)- electrons below the intensity drop-off which do not interact with the sample's surface- and is represented in colour code ranging from black to pale yellow. Four $H_2$ ON cycles are highlighted in white in the reflectivity map, clearly showing that the reflectivity shifts towards higher energies upon hydrogen chemisorption on the surface. (b) The extracted work function $V_S^0$ from the reflectivity map of panel a. The relative work function ($\emptyset_{H_2\ ON}-\emptyset_{NiCoPdW}$) steeply increases as hydrogen is chemisorbs on the surface, reaching a peak after about 3 min of exposure. Black arrows indicate four peaks in the work function over time due to hydrogen exposure (four ON and OFF cycles).